\documentclass[num-refs]{wiley-article}

\usepackage{siunitx}
\usepackage{graphicx}
\usepackage{textcomp}
\usepackage{rotating}
\usepackage{longtable}
\usepackage{amsmath}
\usepackage{afterpage}
\usepackage{multirow}
\usepackage{pifont}
\usepackage{subcaption}
\papertype{Original Article}
\paperfield{Journal Section}

\title{Does Homophily Make Socialbots More Influential?\\ Exploring Infiltration Strategies}

\author[1]{Samaneh Hosseini Moghaddam}
\author[1]{ Mandana Khademi}
\author[1 *]{Maghsoud Abbaspour}

\affil[1]{Faculty of computer science and engineering, Shahid Beheshti Univercity GC, Evin, Tehran, Iran. 1983969411}
\affil[*]{Coresponding Author, Email: maghsoud@sbu.ac.ir }

\corraddress{Maghsoud Abbaspour, Faculty of computer science and engineering, Shahid Beheshti Univercity GC, Evin, Tehran, Iran. 1983969411}
\corremail{maghsoud@sbu.ac.ir}

\runningauthor{Hosseini Moghaddam et al.}

\begin{document}

\maketitle

\begin{abstract}
Socialbots are intelligent software controlling all the behavior of fake accounts in an online social network. They use artificial intelligence techniques to pass themselves off as human social media users. Socialbots exploit user trust to achieve their malicious goals, such as astroturfing, performing Sybil attacks, spamming, and  harvesting private data. The first phase to countermeasure the malicious activities of the socialbots is studying their characteristics and revealing strategies they can employ to successfully infiltrate stealthily into target online social network. In this paper, we investigate the success of using different infiltration strategies in terms of infiltration performance and being stealthy. Every strategy is characterized by socialbots profile and behavioral characteristics. The findings from this study illustrate that assuming a specific taste for the tweets a socialbot retweets and/or likes make it more influential. Furthermore, the experimental results indicate that considering the presence of common characteristics and similarity increase the probability of being followed by other users. This is in complete agreement with homophily concept which is the tendency of individuals to associate and bond with similar others in social networks.  \par

\end{abstract}

\section{Introduction}\label{intro}
Twitter is one of the top five most popular social networks~\cite{Top15Mos49:online}. It facilitates users to connect with each other irrespective of their geographical locations, share their thought expressions, propagate news, promote a celebrity or business brand, and so on. Users can follow other users in order to receive their tweets. It leads to a follower-followee~\footnote{During this paper, in line with the APIs of Twitter, followee is also called \emph{friend}.} relationship. The popularity and the importance role in the world of business and policy, made Twitter an interesting target for attackers. Twitter estimates 23 million of its active users are actually socialbots~\cite{seward2014twitter}.\par
Socialbots are automated softwares which are controlling fake accounts to exploit users trust to achieve their goals. They imitate actual users to pass themselves off as human accounts. Socialbots have verity applications with good or malicious objectives. Some significant malicious applications are astroturfing~\cite{woolley2016automating
}, performing Sybil attacks~\cite{wei2013sybildefender}, spamming~\cite{fire2012strangers, benevenuto2010detecting, lee2010uncovering}, and harvesting private data~\cite{boshmaf2011socialbot}. The first phase to detect or countermeasure malicious socialbots is studying them and revealing strategies they can employ to allure actual users to be successful in terms of social influence and being stealthy. Studying the potential power of characteristics of infiltration strategies is also useful for developing good socialbots.\par
Several researches have been conducted that performed the experiment of socialbot creation to demonstrate the possibility of socialbot network establishment~\cite{boshmaf2013design, elyashar2013homing, zhang2013impact}. Nevertheless, a limited number of researches analyze the characteristics of socialbots to reveal the most infiltrative characteristics~\cite{freitas2015reverse, 7945454}. These studies mainly focused on investigating the impact of basic profile and behavioral related characteristics such as gender, age, following, and so on. Accordingly, there is still considerable ambiguity and unexplored arrangement of characteristics. This paper aims to explore more advanced characteristics to find out the most effective infiltration strategies. To this goal, we use factorial design experiment~\cite{jain1990art} to evaluate the infiltration performance and stealthy of different socialbot strategies, and examine the effectiveness of individual profile and behavioral related characteristics of the socialbots. Consistently using the terminology of the factorial design experiment, all profile and behavioral related characteristics under study assumed to be independent variables which are called factors and the values of the characteristics are called factor levels. Since the factorial design experiment provides more information at a similar or lower cost, it is more efficient than traditional one-factor-at-a-time (OFAT) experiments. Furthermore, the factorial design experiment provides a facility for studying the relationship between different factors. Our experiments are rooted in the hypothesis that there are some relationships between the success in infiltrating into a social network and the profile and behavioral related characteristics of a socialbot. \par
Since there is an exponential relationship between the number of characteristics and the number of strategies come out from the difference of the arrangement of these characteristics, performing a blind study of all characteristics is not feasible. Accordingly, employing some heuristics in choosing the set of characteristics to be studied is advantageous. Thus, in order to cope with the exploding growth in the number of strategies, we employ two heuristics to decide on factors. Additionally, we divide all factors into two groups and do our study in two phases. In the first phase some basic factors including two behavioral characteristics which determine the status of considering or not considering the users' interest, are studied. The interest of a socilabot is determined by the general category of what it retweets and/or likes belongs to. Based on the experimental results, it can be concluded that considering a specific taste for a user is an effective strategy for social infiltration. A primary and raw results of the first phase are reported in~\cite{khademi2017empirical}. A much more detailed analysis is provided in this paper. In the second phase, the focus is on the inspection of the impact of characteristics regarded the homophily along topic of interest which is called the topical homophily ~\cite{kang2012using}. It is speculated that among all profiles and behavioral characteristics studied, the ones which are bound up with topical homophily have a more telling impact on infiltration performance.\par
The rest of the paper is organized as follows. Section 2 briefly surveys related works. Section 3 is dedicated to describing the methodology used to create the socialbots, and the various metrics to quantify the success of them in terms of infiltration performance and being stealthy. In section 4 and 5, a comprehensive analysis of the impact of the various characteristic of two phases of our study in the socialbots’ infiltration performance is presented. Finally, in the section 6, conlusions and directions of future works are provided.\par

\section{Related works}\label{sec:Related Works}
Many researches are dedicated to create and inject a socialbot into an OSNs. This works can be categorized into two main domains: (1) proving the vulnerability of social networks against socialbot infiltration and its implications, (2) investigating the infiltration performance of different socialbot infiltration strategies. \par
The majority of works are aimed at demonstrating the feasibility and implication of socialbots. Boshmaf et al~\cite{boshmaf2013design,boshmaf2011socialbot} studied the use, impact, and implication of socialbots on Facebook. They created a socialbot net consisting of 102 socialbot on Facebook and indicated the vulnerability of OSNs to a large scale infiltration by socialbots. In addition, the feasibility of privacy breach by exploiting socialbots is proved. Elyashar et al~\cite{elyashar2013homing} leveraged socialbots to infiltrate specific users from targeted organization to breach technical information. They show that even security aware users can be duped by socialbots. Zhang et al~\cite{zhang2013impact} constructed a socialbot network consisting of 1000 socialbots on Twitter. They validated the effectiveness and advantage of employing a socialbot network for spamming. Moreover, they demonstrated that socialbots easily can manipulate OSN's repudiations metrics such as Klout score and Twitalyzer. Messias et al~\cite{messias2013you} developed two fake accounts controlled by socialbots on Twitter to verify the possibility of gaining considerable influential scores using very simple strategies. One of the socialbots reached an influence score close to some celebrities. Some other works created and injected socialbots to identify susceptible accounts~\footnote{Users which are alured to follow a socialbot are called susceptible.} and study their distinguished characteristics compared to others~\cite{boshmaf2015integro,aiello2012people}.\par
Some other researchers took a phase further and explore different infiltration strategies to reveal the most significant attributes. Freitas et al~\cite{freitas2015reverse, freitas2016empirical} performed very first experiments in studying the impact of different socialbot strategies on the success of socialbots in terms of  infiltration performance. They created 120 socialbot accounts with different characteristics and behaviors. Investigated strategies were formed by using four characteristics: (1) gender, (2) activity level indicating the volume of posting tweets and following other users, (3) tweet generating strategy including either retweeting or employing automated sentence generators, and (4) the policy of choosing target users. Although 31\% of socialbots were missed during the experiment time period, they showed that activity level and the policy of choosing target users play a significant role. In contrast, gender and generation strategy don’t have much influence on the success of a socialbot. Fazil and his coworker~\cite{7945454} created 98 Twitter socialbots associated to six different countries and studied the influence of some profile characteristics and activities on infiltration performance of a socialbots. They found that socialbots belonging to India are more successful in duping users. Gender has no significant impact and following is the most infiltrative behavior among the activities which is studied. \par
\section{Methodology}
We aim to find out an effective design for socialbots by exploring and evaluating different socialbot infiltration strategies. Each infiltration strategy indicates a set of profile and behavioral related characteristics. The intuition behind our methodology is that creating a socialbot with different characteristics results in different quality in terms of infiltration performance and stealth.\par 
Designing a socialbot infiltration strategy involves decision on a pile of profile and behavioral attributes. Therefore, studying all plausible strategies needs to investigate all feasible arrangements of all profile and behavioral characteristics of a socialbot. There is an exponential relation between the number of investigating characteristics and the number of possible strategies made up of them. Hence, studying all attributes is costly and infeasible.  The number of possible strategies is calculated using equation~\ref{eq:1}, where $A_i$ is the set of distinguished valid values of I-the characteristic and n is the number of investigating characteristics.\par 
\begin{equation}
\Pi _{i=1}^n |A_i|
\label{eq:1}
\end{equation}
To manage the exponentially increasing number of feasible strategies, we focus on some unstudied characteristics which are mainly designed exploiting social theories. Furthermore, we partition all characteristics into two groups and investigate the effect of the characteristics of each group separately. In the first phase, basic profile and behavioral characteristics are studied. During the second phase, some characteristics which are mainly designed on the shade of homophily theory, are investigated while the values of the basic characteristics of the most influential strategy of the first phase are employed. During the first phase, five characteristics which three of them are binary and the two rest have four valid values, are studied. According to equation~\\ref{eq:1}, the number of strategies would be 128. And in second phase, the impact of five binary and one quadratic factors are explored. Thus, the number of strategies would also be 128. Therefore, totally 256 strategies are considered. If all eleven factors of both two phases are studied in one phase, the number of strategies would be 16384, which is 64 times greater than 256. In following subsection some details about creating socialbots and common characteristics and also evaluating strategies, which are identical in both phases are described. 
 
\subsection{Socialbot Creation}
The socialbot creation consists of two main phases: (1) creating a fake account and (2) developing an automated software which controls all activities of that account. All social accounts are created manually. This phase involves deciding on some basic profile characteristics. In addition, developing a software controller includes deciding on behavioral characteristics. Some of these profile and behavioral characteristics are common between all investigated strategies and some are different. We introduce common characteristics in this subsection.\par
For each strategy two socialbots are created which are similar in characteristics which are under studied and different in gender. Twitter user profiles have no attribute indicating the gender of the user. Users usually infer the gender of another user based on the profile picture or the first name attribute of the profile. In our experiment, the names of socialbots are selected form a list of popular Persian names. The profile pictures are chosen from popular photos on the http://hotornot.com/ website or eye-catching pictures of Google search result according to the assigned gender of corresponding socialbot.  \par
In our experiment, all socialbots have the same time zone, which is +3:30 GMT. In addition, the language of all created profiles are set to English and also the language of all reposted tweets are English. As proved in ~\cite{freitas2015reverse} algorithmic sentences can be easily identified. Thus, we don't exploit any automated sentence generation methods like Markov chain to synthetically generate tweets.\par
To inject every new born socialbot into Twitter’s network of users by establishing some primary connections, we made every socialbot tweet a quote and follow ten users which are suggested by Twitter just after creating the fake account. This is an effective strategy to prevent \emph{Twitter’s Trust and Safety team} from suspending our socialbots.   \par
All activities are performed from 9 AM till 12 PM according to indicated time zone and any synchronized behaviors are avoided. Furthermore, we manage periodic behavior as a clue for fake account identification by using random time and random number in a fixed range as the number of times that an activity carry out in a day. Activities are limited to \emph{follow}, \emph{retweet}, and \emph{like}. As it is concluded in~\cite{freitas2015reverse}, the more activity a socialbot do, the more infiltration it achieved. The socialbots with the most high activity level, they introduced, perform 14 actions daily on average. On the other hand, it is conceivable that increasing the volume of activity leads to abnormal behavior and suspending by Twitter defense mechanism. So, we choose the number of activities that every socialbot performs in every day in a range that centered by 14.\par
We create fake accounts manually rather than purchasing them from black markets or using automated profile creation tools, to be able to set profile attributes according to the characteristics in line with strategy assigned to that socialbot and also common attributes mentioned above.\par

\subsection{Strategy Evaluation}\label{strategyEval}
The success of each strategy is evaluated in terms of infiltration performance and being stealthy at the end of the duration of the experiment. We will give score to the strategies by metrics which are measuring infiltration performance quantitatively. However, the stealthy of strategies is evaluated, as a secondary criterion, to determine to what extent they are undetectable. \par
Infiltration performance evaluates how much a socialbot could be effective when employed to share something. As reported in~\cite{cha2010measuring}, \emph{the number of followers} represents the infiltration scale and user’s popularity, but is not related to other important notions of influence such as engaging the audience. Therefore, we use \emph{Klout score} as well. \par
There are two ways to follow a user on Twitter. One is following a user directly, and the other is following a user indirectly by adding she/he to a list. When a user is followed directly, its tweets appear in the ``home feed''. But, if a user is added into a list, its tweets will appear just in that list. When we talk about the number of followers, we mean the total number of direct and indirect followers. If just the number of direct followers is intended, we will mention it explicitly. As it will describe later in following section, some socialbots are followed by some of our other socialbots. This kind of followers is not counted into account when measuring the number of followers metric. \par
Klout score is a popular and standard metric calculated by a private company, Klout.com, which collects information of an account in a social network to evaluate their overall social influence~\cite{rao2015klout}. It is a number between 1 and 100. The more influential a social account is, the higher its Klout Score will be. The exact algorithm used to compute Klout score is not publicly known, but what is admitted is that following measures help to increase Klout score: Followers, retweets, mentions, list memberships.\par
A primary and simple way to evaluate stealthy of a strategy, which is used by some related works~\cite{freitas2015reverse, 7945454}, is whether the socialbot account developed based on the strategy is detected and suspended by Twitter defense mechanism or not. We also use the same criteria to be able to compare our experimental results with theirs. \par

\subsection{Ethical considerations}
In the procedure of both two phases of this study, 256 socialbot accounts were created. The total number of inner and outer links which are established, is about thirty thousand that is ignorable comparing to the size of the Twitter social network. Since all of the tweets reposted by the socialbots are selected from popular tweets which are shared a large number of times, we believe that the nearly forty thousand popular tweets totally shared by the socialbots during 60 days of the experiment, have negligible impact on the ecosystem of Twitter that about 500 million tweets posted daily on it~\cite{Twitter47:online}. However, all socialbot accounts were deleted after gathering all data needed to study them. \par

\section{Phase1: Basic Characteristics}
There are different potentially influential characteristics when different infiltration strategies are explored. Some of them are empirically studied in previous researches and we leveraged the results on forming general characteristics of our socialbots. In the first phase some basic profile and behavioral characteristics are inspected. Profile characteristics include “gender” and “profile picture” which have impact on two remarkably more eye-catching profile attributes: name and profile picture. Aside from two profile characteristics, three behavioral characteristics are leveraged to make different strategies. Detailed description of these characteristics is provided in following subsection. Afterwards, the results of our experiment are presented in separate subsection. \par

\subsection{Socialbot Creation}
The characteristics our study is focused on, during the first phase, are introduced in this subsection. The definition and the different varieties of each characteristic are stated.  \par
1)	\emph{Gender}: Although earlier researches proved that gender has not significant impact on the success of infiltration strategies, we use this characteristic following two goals: (1) to apply gender on first name and profile picture to reach a consistent profile attributes and (2) to circumvent gender bias by generating two different versions of every strategy. \par
2)	\emph{Profile picture}: We exploit two categories of attractive profile pictures. First, the popular portrait photos of http://hotornot.com/ are used. Second, some stunning girlish/boyish photos are selected by googling.  \par
3)	\emph{First target set}: Having some primary follower is essential for the success of subsequent follower acquisition~\cite{boshmaf2011socialbot}. Thus, some earlier researchers made collusion between socialbots and they follow each other at first. Establishing relation between socialbots make them vulnerable to network based fake account detection methods~\cite{yu2006sybilguard, viswanath2010analysis, cao2012aiding, yu2008sybillimit}. To cope with this problem, we divide the activity time period of every socialbot into two consecutive phases. During the first phase, a predefined target users with specific attributes are followed and in the second phase randomly selected users are followed. “First target set” characteristic indicates the specific attributes of users, which are followed in first phase. It could be any of these categories: (1) butterfly accounts, (2) promoter accounts, (3) follow backer accounts, and (4) other socialbots. The butterfly accounts have an extraordinary large number of followers and followings. In addition, the number of their following accounts is greater than 5000~\cite{yang2012analyzing}. The promoter accounts have a large number of followings compared to the number of followers and also have relatively high URL ratio~\cite{yang2012analyzing}. There are Twitter’s accounts which are registered themselves in the known websites and promise to follow back users that follow them. We call this kind of Twitter accounts follow backer accounts. We leverage two web sites to access the list of follow backers: (1) http://www.letsallfollowback.com/, and (2) http://twitterautofollowlist.com/. \par
4)	\emph{User interest}: The intuition behind designing “user interest” is a specific interest assignment to the socialbots to liken them more to the mankind. We consider three different main topics as the types of interest and one level for no specific interests. So four different values for this characteristic are devised: (1) art, (2) sport, (3) politic, and (4) no specific topic. The socialbots exhibit their interests by selecting tweets to repost. For this purpose, socialbots belonging to the first three categories get and repost the most popular tweets by most hot hashtags related to their type of interest. The socialbots of the last category repost most popular tweets regardless of their topic. To choose the most popular hashtags, we use http://hashtagify.me/ web page. This page allows their users to search among more than 70 million Twitter hashtags and find the best ones for their needs based on different criteria such as popularity, relationships, languages, influences and other metrics.  \par
5)	\emph{Liking interesting tweets}: It is a binary characteristic which indicates that whether the socialbot performs like activity or not. The socialbots like tweets under the topic of their interest. \par 
According to the number of varieties for each characteristic and the equation~\ref{eq:1}, the total number of strategies achieved by these characteristics is 128. We create all socialbots on Twitter. They are implemented in Python using Tweepy package. Tweepy is a Python library for accessing Twitter APIs. All socialbots execute using one machine for 40 days that 10 days are dedicated to primary infiltration. We evaluate socialbots in terms of infiltration performance and stealthy just after the end of this time period. Results are provided in next subsection.\par

\subsection{Experimental Results}
As it is stated in subsection~\ref{strategyEval}, the success of socialbots is evaluated using infiltration performance and being stealth. While note of the socialbots created in the first phase is suspended or blocked by Twitter during the 40 days of experiments, we conclude that the socialbots are stealthy enough not to detect by the security mechanisms of Twitter. Table~\ref{table1-stealthiness} compares the duration of experiments, the number of created socialbots, and the number of them which were suspended during the experiments from the preliminary studies and ours. Although the duration of our experiment is longer than others, no one of the socialbots are detected by Twitter defense mechanism.\par

\begin{table*}[b] 
\caption{Comparison of the stealthiness with related works }
\begin{tabular*} {\textwidth} {p{3.6cm} p{3cm} p{2.8cm}p{2.8cm}}
\hline\noalign{\smallskip}
\textbf{Experiment} & \textbf{ Duration of experiment}& \textbf{Number of socialbots} & \textbf{Number of suspended socialbots} \\ 
\noalign{\smallskip}\hline\noalign{\smallskip}
 Freitas et al~\cite{freitas2016empirical} & 30 & 120 & 38\\
Fazil and his coworker~\cite{7945454} & 30 & 98 & 0\\
Phase 1 of our study & 40 & 128 & 0\\
Phase 2 of our study & 40 & 128 & 0\\
\noalign{\smallskip}\hline
\end{tabular*}
\label{table1-stealthiness}
\end{table*}  

In order to decide on how much every characteristic affect social influence of socialbots, infiltration performance of different values of every considered attributes are investigated in terms of \emph{the number of followers} and \emph{the Klout score}. To indicate the density of  the variations clearly and to provide the possibility of easy visual comparison box-plot diagrams are used~\cite{walpole1993probability}.  \par
\begin{figure}[!ht] 
    \centering
  \begin{subfigure}{.5\textwidth}
	\centering
	\includegraphics[width=0.90\linewidth]{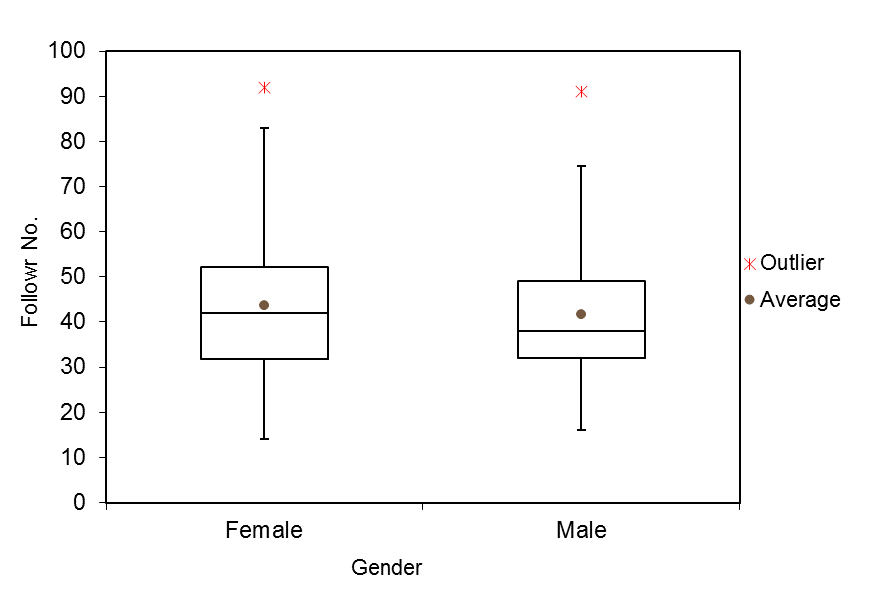}
	\label{gendera}\hfill
	\caption{number of followers}
  \end{subfigure}%
  \begin{subfigure}{.5\textwidth}
	\centering
	\includegraphics[width=0.90\linewidth]{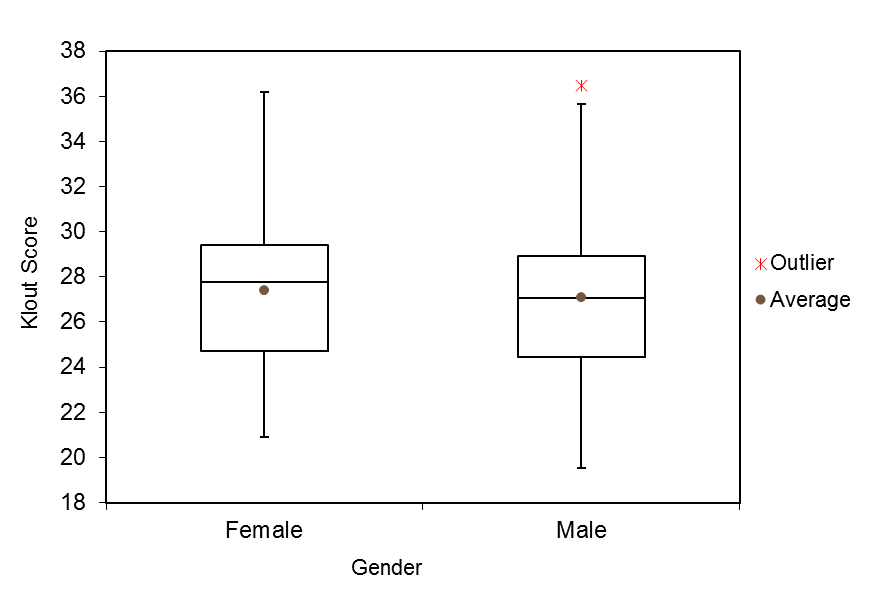}
	\label{genderb}\par
	\caption{Klout score}
  \end{subfigure}%
  \caption{The impact of gender on (a) the number of followers and (b)Klout score}
  \label{fig-gender} 
\end{figure}

Figure~\ref{fig-gender} presents box-plots for the number of followers and Klout score of our socialbots of different gender. It shows that male and female socialbots achieve almost the same number of followers in average and the ranges of Klout score are identical. Female socialbots have little advantage over male socialbots in the maximum number of followers acquired and the median value of Klout score. However, this little advantage is not noticeable as speculated in the previous studies.   \par

\begin{figure}[!ht] 
    \centering
  \begin{subfigure}{.5\textwidth}
	\centering
	\includegraphics[width=0.90\linewidth]{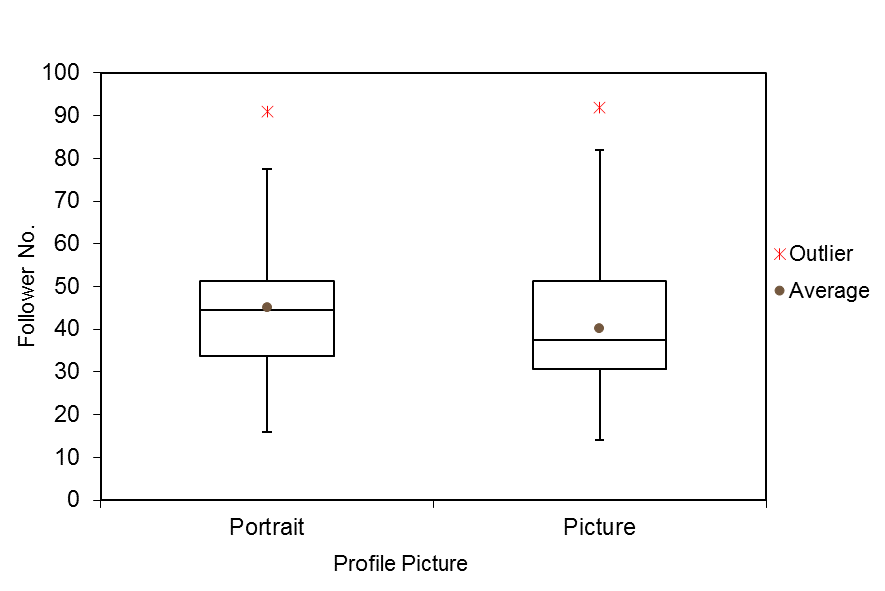}
	\label{fig1a}\hfill
	\caption{number of followers}
  \end{subfigure}%
  \begin{subfigure}{.5\textwidth}
	\centering
	\includegraphics[width=0.90\linewidth]{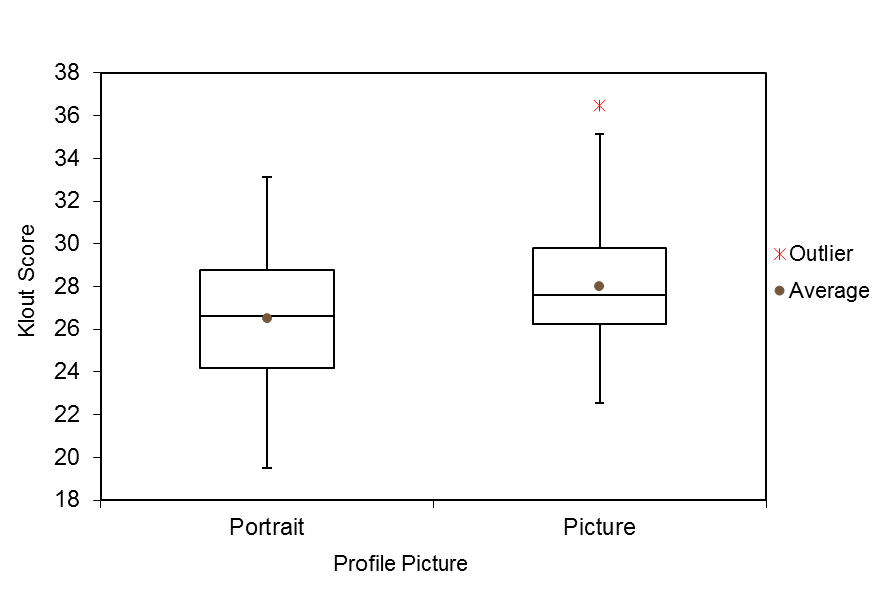}
	\label{fig1b}\par
	\caption{Klout score}
  \end{subfigure}%
  \caption{The impact of profile picture on (a) the number of followers and (b)Klout score}
  \label{fig-picture} 
\end{figure}

In figure~\ref{fig-picture}, two categories of profile pictures, including \textit{portrait} and \textit{eye-catching girlish/boyish picture}, are compared. Although the median number of followers of socialbots with \textit{girlish/boyish profile picture} is lower than socialbots with \textit{portrait} profile picture, \textit{girlish/boyish picture} outdo portrait in Klout score diagram. So the girlish/boyish pictures have slightly more social influence than the portrait pictures.\par 

\begin{figure}[!ht] 
    \centering
  \begin{subfigure}{.5\textwidth}
	\centering
	\includegraphics[width=0.90\linewidth]{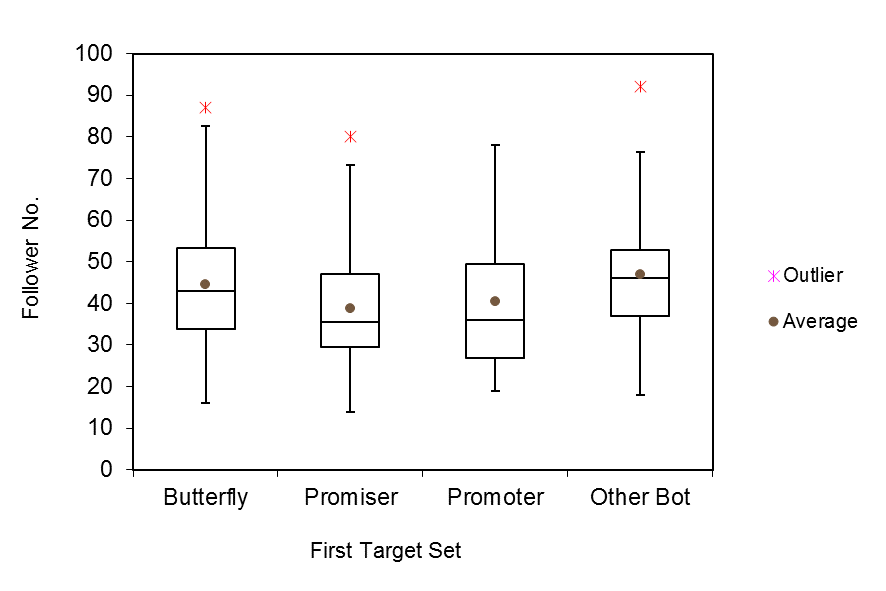}
	\label{fig1a}\hfill
	\caption{number of followers}
  \end{subfigure}%
  \begin{subfigure}{.5\textwidth}
	\centering
	\includegraphics[width=0.90\linewidth]{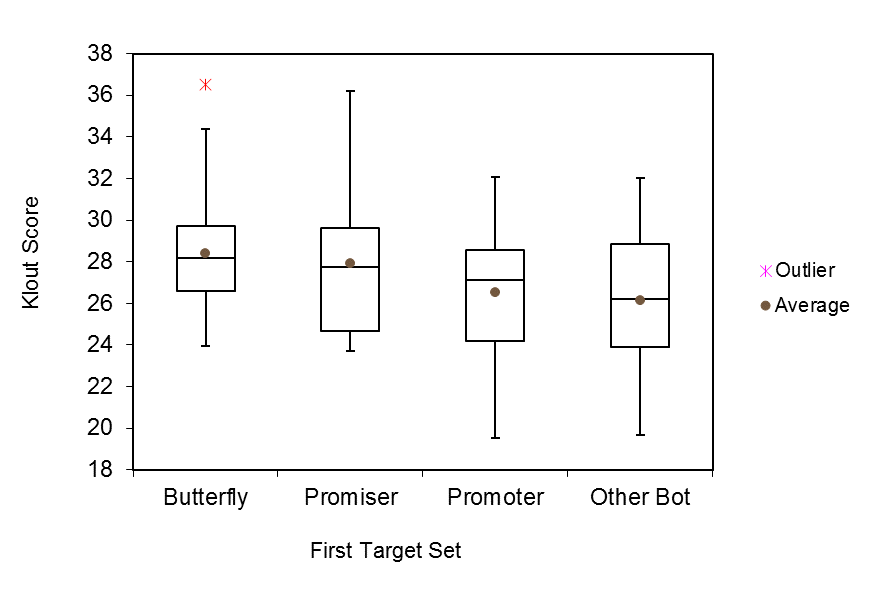}
	\label{fig1b}\par
	\caption{Klout score}
  \end{subfigure}%
  \caption{The impact of first target set on (a) the number of followers and (b)Klout score}
  \label{fig-firstTarget} 
\end{figure}

The infiltration performance of \emph{first target set} characteristic in terms of number of followers and Klout score is illustrated in figure~\ref{fig-firstTarget}. It shows that the range of the number of followers is about the same in all \emph{first target set} categories and little advantage of \emph{other bots} as \emph{first target set} in the median. However, Klout score is a more telling measure for social influence and it shows somewhat superiority of butterflies over other bots regarding both the average and the median of the Klout score distribution. Therefore, targeting butterfly accounts to gain very first social infiltration seems to be a wise choice. \par

\begin{figure}[!ht] 
    \centering
  \begin{subfigure}{.5\textwidth}
	\centering
	\includegraphics[width=0.90\linewidth]{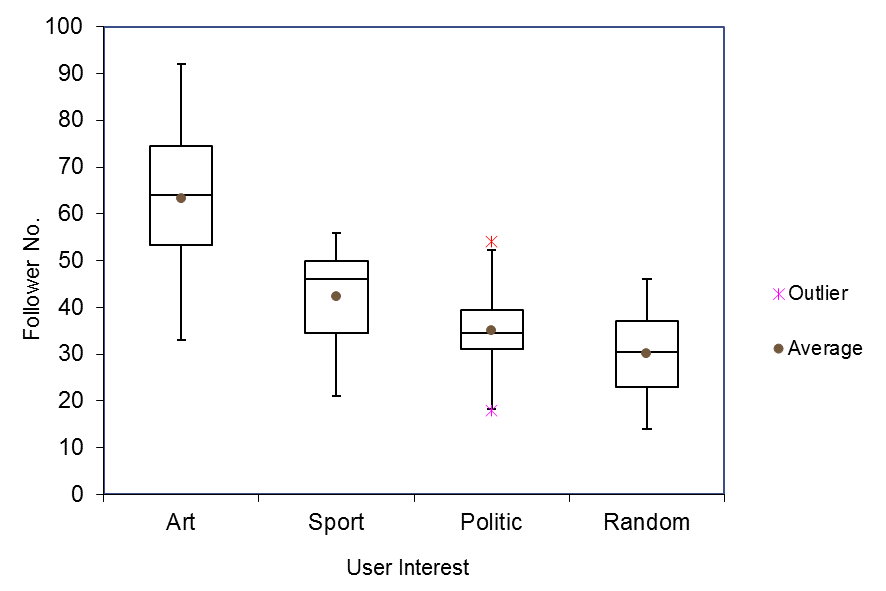}
	\label{fig1a}\hfill
	\caption{number of followers}
  \end{subfigure}%
  \begin{subfigure}{.5\textwidth}
	\centering
	\includegraphics[width=0.90\linewidth]{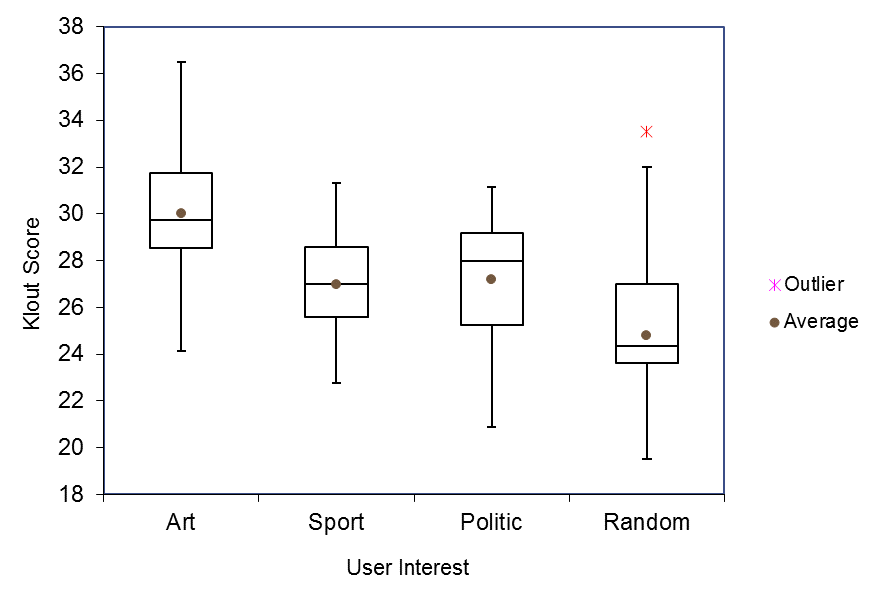}
	\label{fig1b}\par
	\caption{Klout score}
  \end{subfigure}%
  \caption{The impact of user interest on (a) the number of followers and (b)Klout score}
  \label{fig-userInterest} 
\end{figure}

\begin{figure}[!ht]
    \centering
  \begin{subfigure}{.5\textwidth}
	\centering
	\includegraphics[width=0.90\linewidth]{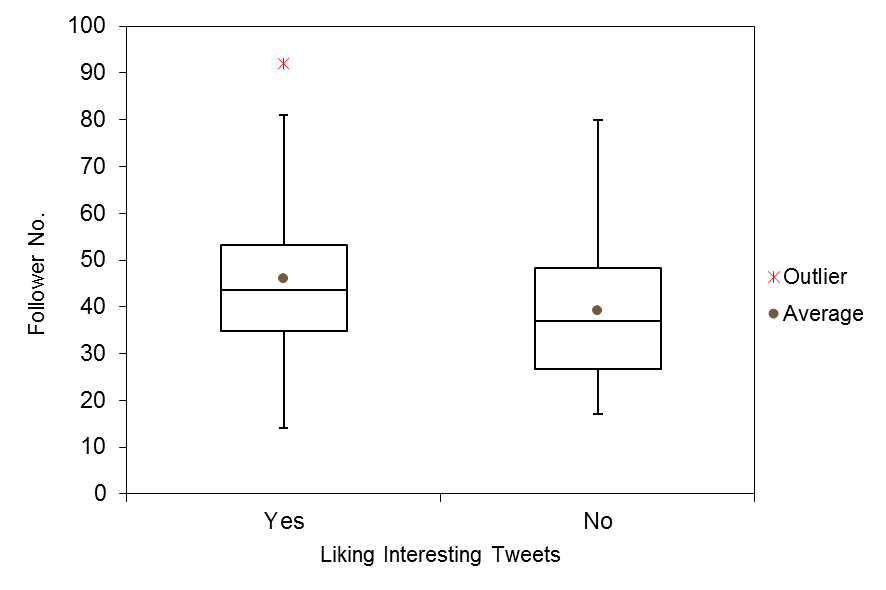}
	\label{fig1a}\hfill
	\caption{number of followers}
  \end{subfigure}%
  \begin{subfigure}{.5\textwidth}
	\centering
	\includegraphics[width=0.90\linewidth]{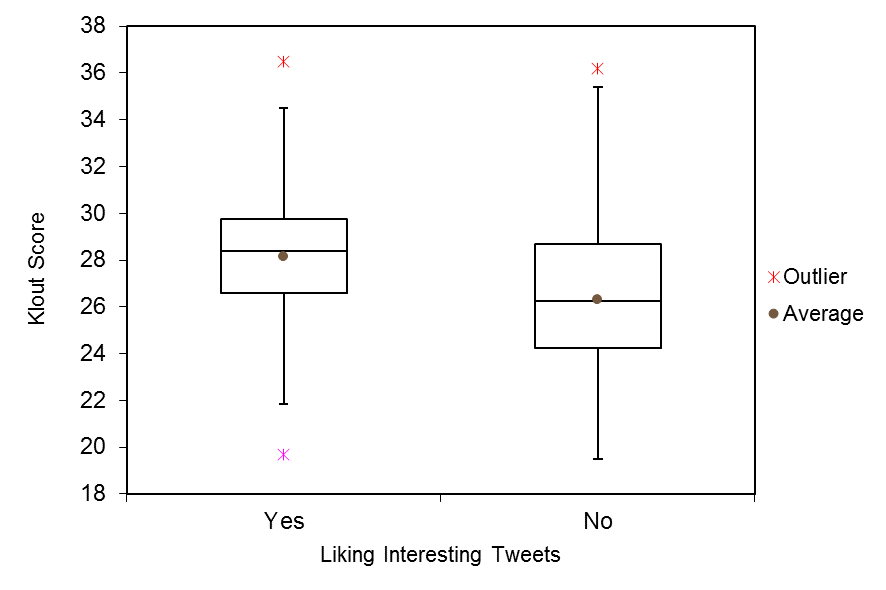}
	\label{fig1b}\par
	\caption{Klout score}
  \end{subfigure}%
  \caption{The impact of liking interesting tweets on (a) the number of followers and (b)Klout score}
  \label{fig-liking} 
\end{figure}

From the figure~\ref{fig-userInterest} we can see a clear correlation between the user interest and both of the infiltration performance measures. Hence, being interested in tweets related to art boosts social influence. It is noteworthy that considering a specific topic as user interest against random tweet increase the probability of success. Liking tweets under the topic of user interest put emphasis on the user's taste. As shown in figure~\ref{fig-liking}, exhibiting user interest by liking tweets have a positive effect on infiltration performance. Furthermore, the variability of both the number of followers and Klout score is less in socialbots liking interesting tweets and thus it means that it is more reliable and there is more probability for socialbot to promote its social influence.   \par

As it is predictable based on the above analysis, the more successful socialbot of our experiment is the one which is male, has girlish/boyish picture, followers butterfly accounts as first target set, is interested in art, and likes tweets related to its interest. The only characteristic which the most successful socialbot has and it is not the dominant one, is its gender. As it is considered former, gender doesn't have a noticeable impact on the infiltration performance. \par
In order to study the contribution of each characteristic in gaining social influence, we conduct a sensitivity analysis. We employ sensitivity index measure to assess the parameter sensitivity. Sensitivity index calculates the normalized difference of output while one input parameter changing from its minimum value to its maximum value, in order to quantify the influence of that input parameter on the output~\cite{hamby1994review}. Sensitivity index is formulated in Equation~\ref{eq:12}, where $D_{min}$ and $D_{max}$ represent the minimum and maximum output values, respectively, while varying the input parameter from its minimum value to its maximum value. \par

 \begin{equation}
SI = \frac{D_{max} -D_{min}}{D_{max}}
\label{eq:12}
\end{equation}

\begin{figure}[!ht]
	\centering
	\includegraphics[width=3.3in]{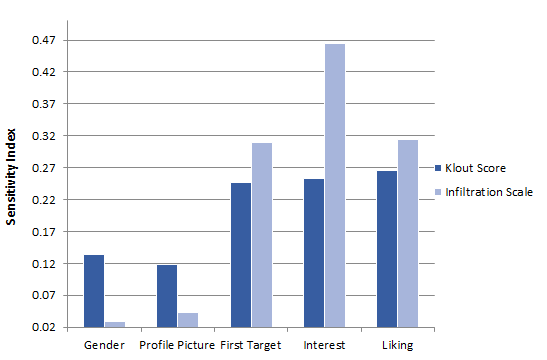}
	\caption{The sensitivity index of each basic parameter takes values between 0 and 1. The higher is the sensitivity
	index the more influent is the parameter on the Klout score.}
	\label{fig-run1SI} 
\end{figure}

In case the input parameters are nominal, the minimum and maximum values of output are considered while the input parameter changes in nominal values. Figure~\ref{fig-run1SI} demonstrates the sensitivity index of basic characteristics. The user interest and liking interesting tweets have the most impact on infiltration scale and Klout score, respectively. As speculated, \emph{User interest} and \emph{liking interesting tweets} which  bound up with user's taste, have more considerable impact on infiltration performance than other characteristics.\par

\par

\begin{figure}[!ht]
    \centering
    \includegraphics[width=3.3in]{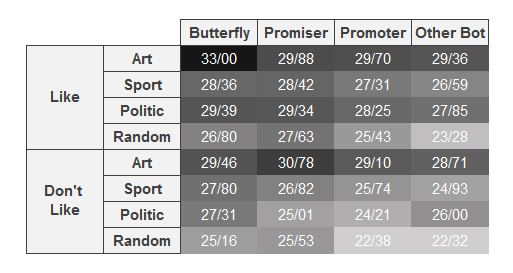}
    \caption{The average value of Klout score for different value of the most influential parameters: Interest, liking, and first target set.}
    \label{heat1}
\end{figure}

In order to study the interaction among three most effective parameters, the average value of Klout score of socialbots with the specified values of these effective parameters are displayed as a heat map diagram in figure~\ref{heat1}. As it can be perceived that the effect of one of those parameters on the Klout score is not constant and the effect differs at different values of the others. For example, the differences between the average of the Klout score of socialbots which are interested in art tweets and like art tweets and the average of the Klout score of socialbots which are interested in sport tweets and like sport tweets varies for different \textit{first target set}. Precisely, when the first target set is set to \emph{butterfly}, the effect of being interested in art on the value of Klout score in comparison to sport, is greater than when the first target set is set to \emph{promiser} or \emph{promoter}. In like manner, the effect of the value of every three most effective parameters varies, depending on the value of others. \par

\par

\section{Phase2: Advanced Characteristics}
Homophily is a social theory which explains friendship creation process. The theory explain that the individuals in a social network tend to become friends with someone who is similar to them. The more similarity, the more likelihood to become friends. Many researches have studied homophily along demographic dimensions such as gender, age, location, and language~\cite{mcpherson2001birds, reagans2004make, yuan2006homophily}. Besides demographic similarity, shared interests, followees, and followers are also influential factors that are helpful in finding similar friends as well~\cite{carullo2015triadic}. Different kinds of similarity arrange different forms of homophily. The similarity in information, taste and attitudes is the origin of hidden value homophily~\cite{mcpherson2001birds}. In addition, triadic closure theory states that if two individuals in a social network have mutual friends, then the probability that they become friends in the future will increase. If having common friends is considered as a dimension of homophily, triadic closure is a special case of homophily theory which is called structural homophily~\cite{robins2009closure}. \par
Inspiring from homophily theory, in this phase, we design some new characteristics and validate the impact of them on the infiltration performance. Notice that common strategies between all socialbots explained former, including (1) follow only English users and (2) retweet and like just English tweets, are aligned with the increasing demographic homophily. \par

\subsection{Socialbot Creation}
In the second phase of our empirical study, all of the socialbots resemble the characteristics of the most infiltrative socialbot of the first phase which is the one which has girlish/boyish picture, follow butterfly accounts as first target set, is interested in art, and likes tweets related to its interest. As it is stated former, gender doesn't have a noticeable impact on the infiltration performance. However, to avoid gender bias, for every strategy produced by different levels of the following factors, two socialbots with different gender are created. The factors our study is focused on, during the second phase, are presented as follows: \par
1) \emph{Target users}: Four levels are considered for the selection of target users to be followed by the socialbots, regarding structural homophily and triadic closure, including the followers of its followers, the followers of its friends, the followers of its neighbors, and randomly selected users. Whereas the term neighbor refers to any type of follower or followee relationship.\par
2) \emph{Retweet by hashtag}: It is a binary characteristic which determines if the socialbot increase the homophily between the user which it follows and itself, or not. If the value of this characteristic is yes, it means that the socialbot will search for a popular tweet which has at least one common hashtag with the latest hashtaged tweet of followed user, if exists. A tweet is counted as a popular tweet if it has at least one hashtag which is included in the 14 most popular hashtag in  http://hashtagify.me/ web page.  \par
3) \emph{Retweet}: In the similar manner as \emph{retweet by hashtag}, it is a binary factor which if it is set to yes, when the socialbot follow an account it retweet one of its recent tweets shortly afterwards. If this factor is set to no, the socialbot do not retweet the tweets of its friends. This characteristic controls an aspect of homophily in interest and taste.\par
4) \emph{Like}: Likewise, as explained above about previous characteristic, this characteristic is also binary and indicates whether the socialbot like a tweet shared by an account, subsequent to following it. In the same way, it focuses on an aspect of hidden value homophily.   \par
5) \emph{Considering recent activities}: In order to study the impact of considering the presence of recent activity of users which are chosen to be followed, in success of a socialbot, two different policies are taken into account: (1)follow solely active users, (2)follow randomly regardless of its activity state.\par
According to the equation~\ref{eq:1}, the total number of strategies achieved by the characteristics of phase 2 is 128. We create all the socialbot accounts manually on Twitter. The socialbots are implemented in Python using Tweepy package. All socialbots execute using one machine for 40 days that 10 days are dedicated to primary infiltration. We evaluate socialbots in terms of infiltration performance and stealthy just after this time period. Results are provided in next subsection.\par

\subsection{Experimental Results}
Over the duration of the second phase of our experiment, the socialbots received in total 27537 follows from 24845 distinct users which is about triple the total number of followers of the socialbots of the first phase. The main concentration of this phase is the evaluation of the impact of employing homophily based characteristics on the success of a socialbot. Table~\ref{table:cmp}  displays the average of Klout score and the follower number for socialbots in four following categories: (1) socialbots which employ no homophily based characteristic, (2) socialbots exploit hidden value homophily based characteristics, including \emph{Retweet by hashtag}, \emph{Retweet}, and \emph{Like}, (3) socialbots which make use of structural homophily based characteristic, namely \emph{Target users}, and (4) socialbots which apply both hidden value and structural homophily based characteristics. It can be seen from the summary statistics provided in table~\ref{table:cmp} that although employing any of hidden value and structural homophily results in a better outcome than using no homophily based strategy, using both of them simultaneously significantly improve the social influence and the popularity of socialbots corresponding to Klout and the number of followers, respectively. These results provide support for the hypothesis that homophily based characteristics make socialbots more influential. \par

\begin{table*}[t] 
\caption{Comparison of the infiltration performance of homophily policies }
\begin{tabular*} {\textwidth} {p{7cm} p{3cm} p{3cm}}
\hline\noalign{\smallskip}
\textbf{Category} & \textbf{ Klout Score} & \textbf{Follower No.} \\ 
\noalign{\smallskip}\hline\noalign{\smallskip}
No Homophily & 27 & 80\\
Hidden Value Homophily & 30 & 80\\
Structural Homophily & 31 & 136\\
Both & \bfseries 35 & \bfseries 209\\
\noalign{\smallskip}\hline
\end{tabular*}
\label{table:cmp}
\end{table*}    

As it is reported in table~\ref{table1-stealthiness}, none of the socialbots created in phase 2 is suspended. Thus, the stealthiness of the strategies employed during this phase is acceptable. In order to consider the impact of each individual characteristic on gaining popularity and influence in the Twitter, the number of followers and Klout score are investigated respectively. These two criteria are measured at the end of the duration of the experiment to quantify how successful a socialbot is. \par
\begin{figure}[!ht] 
    \centering
  \begin{subfigure}{.5\textwidth}
	\centering
	\includegraphics[width=0.90\linewidth]{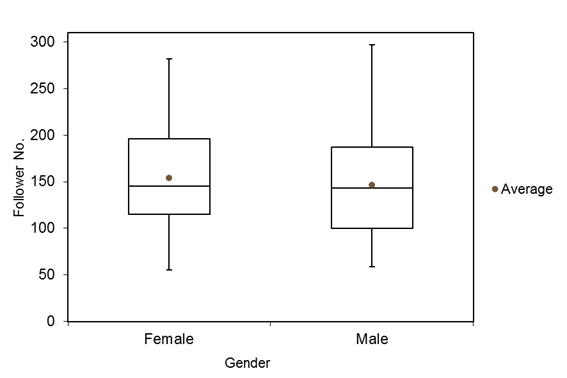}
	\label{gendera}\hfill
	\caption{number of followers}
  \end{subfigure}%
  \begin{subfigure}{.5\textwidth}
	\centering
	\includegraphics[width=0.90\linewidth]{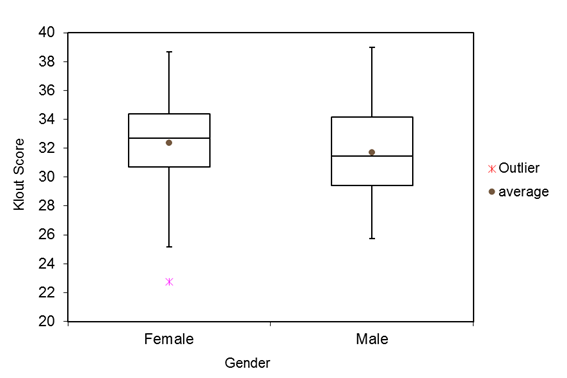}
	\label{genderb}\par
	\caption{Klout score}
  \end{subfigure}%
  \caption{The impact of gender on (a) the number of followers and (b)Klout score}
  \label{fig-gender-2} 
\end{figure}

Although gender is not one of the characteristics studied in this phase, to reconfirm our results in phase one, we start by investigating the impact of the gender. Figure~\ref{fig-gender-2} illustrates box plots that summarize the distributions for the number of followers and the Klout score of socialbots by gender. Although gender has no noticeable impact on the median follower number, the average of the number of followers for female socialbots is slightly more than male ones. Klout score indicates a kind of more success for females. However, our Results confirm with previous findings~\cite{freitas2015reverse, freitas2016empirical,7945454} stating that gender has no significant impact on infiltration performance. \par

\begin{figure}[!ht] 
    \centering
  \begin{subfigure}{.5\textwidth}
	\centering
	\includegraphics[width=0.90\linewidth]{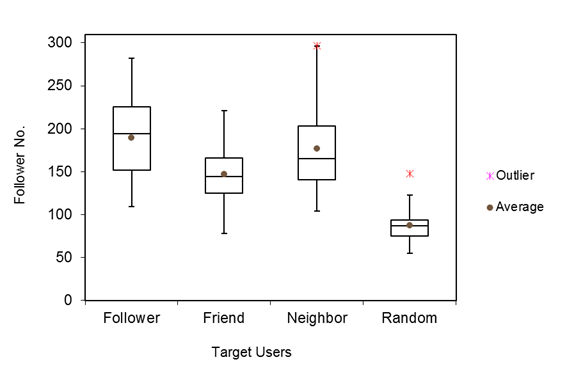}
	\label{targetUsera}\hfill
	\caption{number of followers}
  \end{subfigure}%
  \begin{subfigure}{.5\textwidth}
	\centering
	\includegraphics[width=0.90\linewidth]{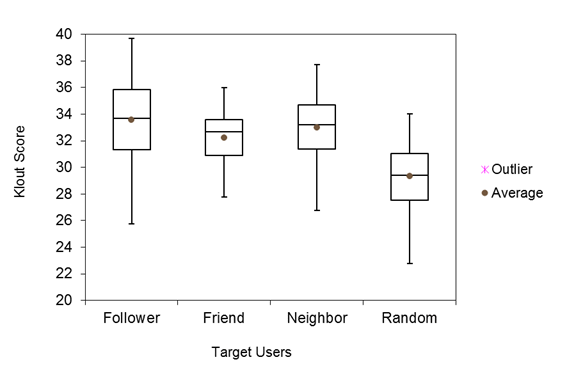}
	\label{targetUserb}\par
	\caption{Klout score}
  \end{subfigure}%
  \caption{The impact of target users on (a) the number of followers and (b)Klout score}
  \label{fig-targetUser} 
\end{figure}

As described former, \emph{target users} is a characteristic  which indicates the policy of following other users. Four different levels are defined for this characteristic, including the follower of its followers, the followers of its friends, the followers of its neighbors, and randomly selected user. Figure~\ref{fig-targetUser}a and b shows the quartiles of the distribution of the number of followers and Klout score, respectively, for each target user. From the figure, it is evident that there is a significant difference in the popularity of randomly following users or following the followers of users in the neighborhood. There is a similar pattern in the case of Klout score. Overall, these results reveal that infiltration strategies employing structural homophily based characteristics are better choices. The plot boxes in the figure~\ref{fig-targetUser} suggest that among structural homophily based policies studied following the followers of the followers of the socialbot results in moderately more successful infiltration strategy in terms of either the number of followers or Klout score. Hence, it could conceivably be hypothesized that the users are more confident in their friends than their followers. Thus, the probability of following back a new follower which is a friend of friend is more than the probability of following back a new follower which is a friend of followers. \par  

\begin{figure}[!ht] 
    \centering
  \begin{subfigure}{.5\textwidth}
	\centering
	\includegraphics[width=0.90\linewidth]{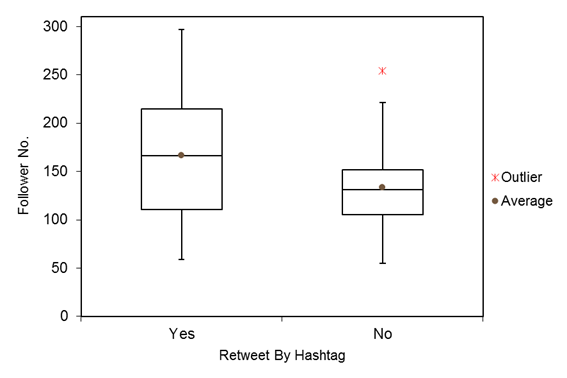}
	\label{retweetBtHasha}\hfill
	\caption{number of followers}
  \end{subfigure}%
  \begin{subfigure}{.5\textwidth}
	\centering
	\includegraphics[width=0.90\linewidth]{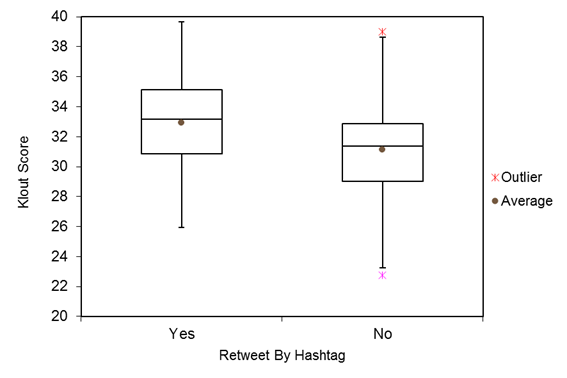}
	\label{retweetBtHashb}\par
	\caption{Klout score}
  \end{subfigure}%
  \caption{The impact of retweet by hashtag on (a) the number of followers and (b)Klout score}
  \label{fig-retweetBtHash} 
\end{figure}

As stated earlier, half of our socialbots in phase 2 after following a user, select a hashtag of one of its recently posted tweets, if exists, and search for a popular tweet having the same hashtag, whereas the other half do not repost such a popular tweet with common hashtag. This binary factor which is called \emph{retweet by hashtag} inspired by the hidden value homophily. Figure~\ref{fig-retweetBtHash} shows the main descriptive statistics of the distribution of the number of followers and Klout score, respectively, for two levels of this binary factor in the form of box plot. As can be seen from this figure, reposting a popular tweet by common hashtag brings slightly better results. Two discrete reasons emerged from this. First, as it is assumed, using homophily based characteristics in the design of the infiltration strategy improves the probability of follower achievement and as a result, boosts social influence. Second, embedding this characteristic leads to higher activity level which is confirmed that it results in better infiltration performance~\cite{freitas2016empirical}.\par

\begin{figure}[!ht] 
    \centering
  \begin{subfigure}{.5\textwidth}
	\centering
	\includegraphics[width=0.90\linewidth]{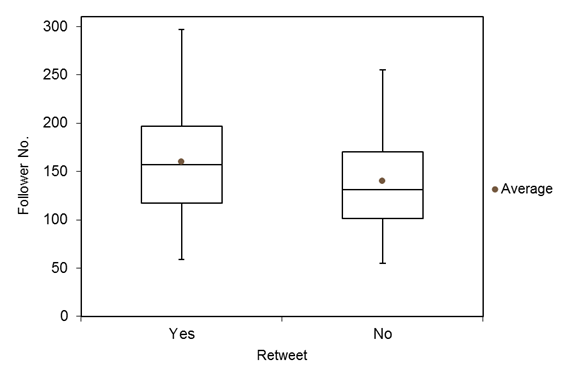}
	\label{retweeta}\hfill
	\caption{number of followers}
  \end{subfigure}%
  \begin{subfigure}{.5\textwidth}
	\centering
	\includegraphics[width=0.90\linewidth]{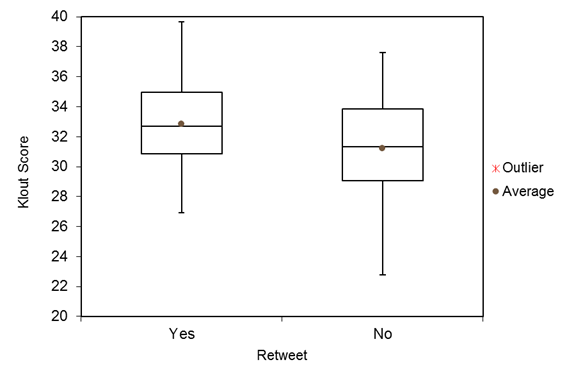}
	\label{retweetb}\par
	\caption{Klout score}
  \end{subfigure}%
  \caption{The impact of retweet on (a) the number of followers and (b)Klout score}
  \label{fig-retweet} 
\end{figure}

Recall from previous subsection half of our socialbots repost one of the recently posted tweets of a user after following it, while the other half does not do that. This characteristic denoted as \emph{retweet}. Figure~\ref{fig-retweet} compares the summary statistics of the number of followers and Klout score for two levels of this binary characteristic in the form of box plot. It shows that the median follower number and Klout score of the socialbots which retweet after following are higher. In the same way as \emph{retweet by hashtag}, it is therefore probable that the superiority of retweet after following is because of either homophily enhancement or activity increase.\par

\begin{figure}[!ht] 
    \centering
  \begin{subfigure}{.5\textwidth}
	\centering
	\includegraphics[width=0.90\linewidth]{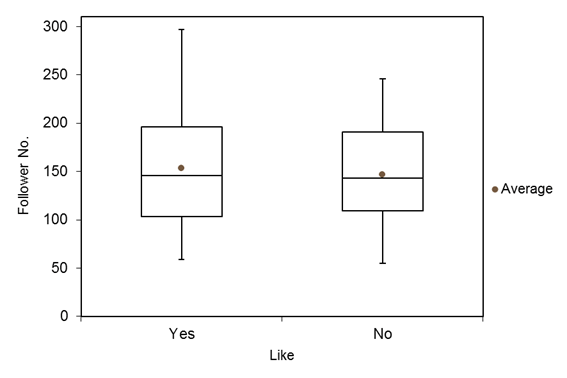}
	\label{likea}\hfill
	\caption{number of followers}
  \end{subfigure}%
  \begin{subfigure}{.5\textwidth}
	\centering
	\includegraphics[width=0.90\linewidth]{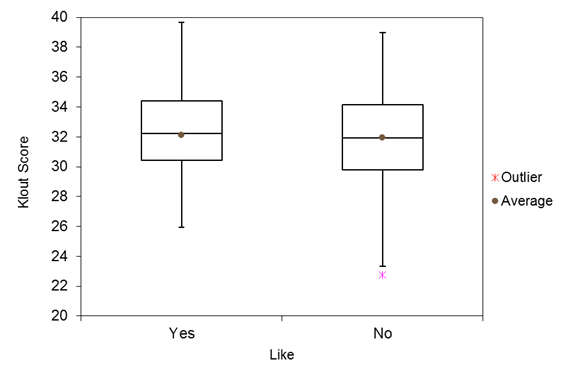}
	\label{likeb}\par
	\caption{Klout score}
  \end{subfigure}%
  \caption{The impact of like on (a) the number of followers and (b)Klout score}
  \label{fig-like} 
\end{figure}

Another factor suggested to study the beneficence of homophily is \emph{like after following}. Respecting to this characteristic, half of our socialbots like one of the recently posted tweets of a user after following it, while the other half don not like. As it is illustrated in figure~\ref{fig-like}, although there is no significant difference in median popularity acquired by socialbots of different levels of this factor, since the upper whisker is longer in case of like after following, the average of the follower number is more than the other. The median and inter-quartile range of both two groups are roughly the same for Klout score. However, the average of exploiting this strategy is slightly more, because of having longer upper whisker and shorter lower whisker than the other.\par

\begin{figure}[!ht] 
    \centering
  \begin{subfigure}{.5\textwidth}
	\centering
	\includegraphics[width=0.90\linewidth]{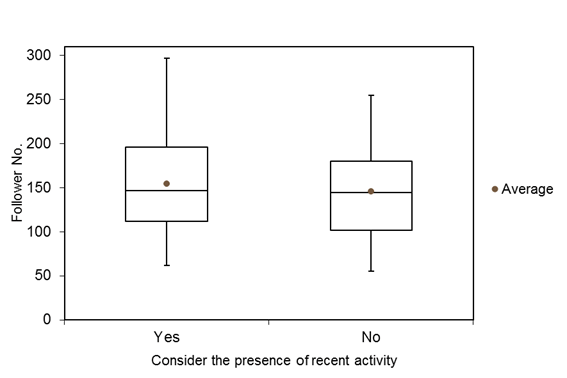}
	\label{considerActivitya}\hfill
	\caption{number of followers}
  \end{subfigure}%
  \begin{subfigure}{.5\textwidth}
	\centering
	\includegraphics[width=0.90\linewidth]{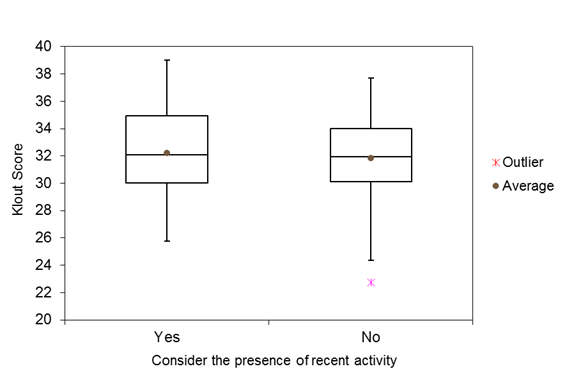}
	\label{considerActivityb}\par
	\caption{Klout score}
  \end{subfigure}%
  \caption{The impact of considering the presence of recent activity on (a) the number of followers and (b)Klout score}
  \label{fig-considerActivity} 
\end{figure}

The last characteristic is the one deciding whether consider the presence of recent activity of who is followed or not. The results are demonstrated in figure~\ref{fig-considerActivity}. Surprisingly, the median and the average of Klout score is similar for both of two state of this characteristic. Although the median and the average of the number of  followers show a bit better results for considering recent activity, the difference is not significant. \par

\begin{figure}[!ht]
	\centering
	\includegraphics[width=3.3in]{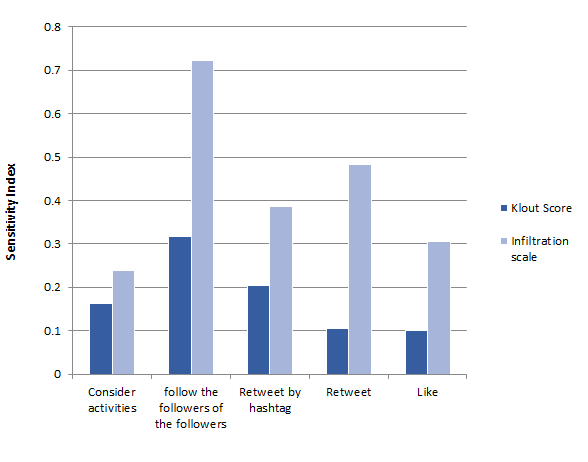}
	\caption{The sensitivity index of each advanced parameter takes values between 0 and 1. The higher is the sensitivity
	index the more influent is the parameter on the Klout score.}
	\label{fig-run2SI} 
\end{figure}

The more successful socialbot of the second phase is the one which is female, her target user set consists of the followers of her followers, retweet similar tweets by hashtag, retweet a randomly selected tweet of who followed, and like a randomly selected tweet of who followed. As it is considered former, gender doesn't have a noticeable impact on the infiltration performance. In order to study the contribution of other characteristics in gaining social influence, we conduct a sensitivity analysis. Figure~\ref{fig-run2SI} demonstrates the sensitivity index of advanced characteristics. The policy of selecting the target users has the most impact on infiltration scale and Klout score. Since the characteristics studied in the second phase mainly designed inspiring homophily theory which focused on the mechanism of the formation of relations, they have more impact on the infiltration scale than Klout score.\par

\par

%

\section{Conclusion}
This study set out to determine the influence of different characteristics on the success of the infiltration strategy of socialbots. For this purpose ten characteristics are arranged to be investigated during two phases. The first phase is managed to investigate some basic profile and behavioral characteristics. It has shown that assigning a specific taste to a socialbot make it more popular and also further influential. Five other characteristics, mainly inspired by the homophily principle, are studied during phase two. Among all profiles and behavioral characteristics studied, characteristics which are bound up with homophily principle have a more telling impact on infiltration performance. As the characteristics studied in the second phase mainly designed inspiring homophily theory which focused on the mechanism of the formation of relations, they have more impact on the infiltration scale than Klout score.\par
Further experimental investigations are needed to find out more influential characteristics on infiltration strategy design. Since there is an exponential relation between the number of exploring factors and the number of possible strategies employing these factors, it is recommended to candidate influential characteristics to explore using social principles. Another possible area of future research would be to investigate the correlation between characteristics.\par

\bibliography{references}
%
%

\end{document}